\documentclass[preprint,aps]{revtex4}
\usepackage[]{amsmath}
\usepackage{graphicx}

\begin{document}


%
%

\title{Nonlocal chiral quark models with Polyakov loop at finite
temperature and chemical potential}

\author{G.A.Contrera$^{a,b}$,
M. Orsaria$^{a,b}$ and
N.N. Scoccola$^{a,b,c}$\\}
\address{
$^{a}$ Physics Department, Comisi\'on Nacional de Energ\'{\i}a At\'omica, Av.Libertador 8250, 1429 Buenos Aires,
Argentina\\
$^{b}$ CONICET, Rivadavia 1917, 1033 Buenos Aires, Argentina\\
$^{c}$ Universidad Favaloro, Sol{\'\i}s 453, 1078 Buenos Aires, Argentina}


\begin{abstract}
We analyze the chiral restoration and deconfinement transitions in
the framework of a non-local chiral quark model which includes terms
leading to the quark wave function renormalization, and takes care
of the effect of gauge interactions by coupling the quarks with
the Polyakov loop. Non-local interactions are described by considering
both a set of exponential form factors, and a set of form factors obtained
from a fit to the mass and renormalization functions obtained in lattice
calculations.
\end{abstract}
\maketitle


\section{Introduction}

The detailed understanding of the behavior of strongly interacting matter
under extreme conditions of temperature and/or density has become an issue
of great interest in recent years. Unfortunately, even if a significant progress
has been made on the development of ab initio calculations such as lattice QCD, these
are not yet able to provide a full understanding of the QCD phase diagram and the related
hadron properties, due to the well-known difficulties of dealing with small
current quark masses and finite chemical potentials. Thus it is important
to develop effective models that show consistency with lattice results and
can be extrapolated into regions not accessible by lattice calculation
techniques. Recently, models in which quark fields interact via local four point
vertices and where the Polyakov loop is introduced to account for the confinement-deconfinement
phase transition (so-called Polyakov-Nambu-Jona-Lasinio (PNJL) models
\cite{Meisinger:1995,Fukushima:2003,Megias:2004,Ratti:2005,Roessner:2006})
have received considerable attention.
Here, we consider a non-local extension of these PNJL models, which includes
terms leading to the quark wave function renormalization. Two different
parameterizations are used: an exponential form, and a parametrization based
on a fit to the mass and renormalization function obtained in lattice
calculations. In the context of this type of model the properties of the
vacuum and meson sectors at $T=\mu=0$ have been studied in
Ref.\cite{Noguera:2008}.

This contribution is organized as follows. In Sec. 2 we introduce
the model lagrangian and its parameterizations. In Sec. 3 we present
and discuss our results for the behavior of some thermodynamical
properties and the corresponding phase diagrams. Finally, in Sec. 4
our main conclusions are summarized.

\section{The model and its parametrizations}

We consider here a nonlocal SU(2) chiral quark model which
includes quark couplings to the color gauge fields. The
corresponding Euclidean effective action is given by
\begin{eqnarray}
\! \! \! \! \! \! \! \!
 S_{E}\! &\! =\! & \! \! \int \! d^{4}x \left\{
\bar{\psi}(x) [ -i \rlap/D+\hat{m} ]
\psi(x) \! - \! \frac{G_{S}}{2}
\Big[ j_{a}(x)j_{a}(x)\! - \!  j_{P}(x) j_{P}(x) \Big]
\! + \!
{\cal U}\,(\Phi[A(x)])\right\} ,
\label{action}
\end{eqnarray}
where $\psi$ is the $N_{f}=2$ fermion doublet $\psi\equiv(u,d)^T$,
and $\hat{m}=diag(m_{u},m_{d})$ is the current quark
mass matrix, in what follows we consider isospin symmetry, that is
$m_{f}=m_{u}=m_{d}$. The fermion kinetic term includes a covariant
derivative $D_\mu\equiv \partial_\mu - iA_\mu$, where $A_\mu$ are
color gauge fields. The nonlocal currents
$j_{a}(x),j_{P}(x)$ are given by
\begin{align}
j_{a}(x)  &  =\int d^{4}z\ g(z)\ \bar{\psi}\left(
x+\frac{z}{2}\right)
\ \Gamma_{a}\ \psi\left(  x-\frac{z}{2}\right)  \ ,
\nonumber\\
j_{P}(x)  &  =\int d^{4}z\ f(z)\ \bar{\psi}\left(
x+\frac{z}{2}\right) \ \frac{i
{\overleftrightarrow{\rlap/\partial}}}{2\ \kappa_{p}}
\ \psi\left(  x-\frac{z}{2}\right) \ ,  \label{currents}%
\end{align}

Here,
$\Gamma_{a}=(\leavevmode\hbox{\small1\kern-3.8pt\normalsize1},i\gamma
_{5}\vec{\tau})$ and $u(x^{\prime}){\overleftrightarrow{\partial}%
}v(x)=u(x^{\prime})\partial_{x}v(x)-\partial_{x^{\prime}}u(x^{\prime})v(x)$.
The functions $g(z)$ and $f(z)$ in Eq.(\ref{currents}), are
nonlocal covariant form factors characterizing the corresponding
interactions. The scalar-isoscalar component of the $j_{a}(x)$
current will generate the momentum dependent quark mass in the
quark propagator, while the ``momentum'' current, $j_{P}(x),$ will
be responsible for a momentum dependent wave function
renormalization of this propagator.

To proceed it is convenient to perform a standard bosonization of
the theory. Thus, we introduce the bosonic fields
$\sigma_{1,2}(x)$ and $\pi_a(x)$, and integrate out the quark
fields. In what follows, we work within the mean-field
approximation (MFA), in which these bosonic fields are replaced by
their vacuum expectation values $\bar \sigma_{1,2}$ and  $\bar
\pi_a = 0$. Next, we extend the so obtained bosonized effective MFA
action to finite temperature $T$ and chemical potential $\mu$
using the Matsubara formalism. Concerning the gluon fields we will
assume that they provide a constant background color field $A_4 =
i A_0 = i g\,\delta_{\mu 0}\, G^\mu_a \lambda^a/2$, where
$G^\mu_a$ are the SU(3) color gauge fields. Then the traced
Polyakov loop, which is taken as order parameter of confinement,
is given by $\Phi=\frac{1}{3} {\rm Tr}\, \exp( i\beta \phi)$,
where $\beta = 1/T$, $\phi = i A_0$. We will work in the so-called
Polyakov gauge, in which the matrix $\phi$ is given a diagonal
representation $\phi = \phi_3 \lambda_3 + \phi_8 \lambda_8$, which
leaves only two independent variables, $\phi_3$ and $\phi_8$.
Owing to the charge conjugation properties of the QCD
Lagrangian~\cite{Dumitru:2005}, the mean field value of the
Polyakov loop field $\bar \Phi$ is expected to be a real quantity.
In addition, we assume as usual that $\phi_3$ and $\phi_8$ are
real-valued fields~\cite{Roessner:2006}, this implies that $\phi_8
= 0$, then $\bar{\Phi} = [ 2 \cos(\phi_3/T) + 1 ]/3$.

Within this framework the mean field thermodynamical potential
$\Omega^{\rm MFA}$ results
\begin{align}
\Omega^{\rm MFA} =  \,- \,\frac{4 T}{\pi^2} \sum_{c} \int_{p,n}
\mbox{ln} \left[ \frac{ (\rho_{n, \vec{p}}^c)^2 +
M^2(\rho_{n,\vec{p}}^c)}{Z^2(\rho_{n, \vec{p}}^c)}\right]+
\frac{\bar \sigma_1^2}{2\,G_S} + \frac{\kappa_p^2\ \bar
\sigma_2^2}{2\,G_S} + {\cal{U}}(\bar \Phi ,T) \ , \label{granp_reg}
\end{align}
Here, the shorthand notation $\int_{p,n} = \sum_n \int d^3\vec
p/(2\pi)^3$ has been used, and $M(p)$ and $Z(p)$ are given by
\begin{eqnarray}
M(p)  =  Z(p) \left[ m_{f} + \bar{\sigma}_{1} \ g(p) \right]
\qquad , \qquad
Z(p)  =  \left[ 1 - \bar{\sigma}_{2} \ f(p) \right]^{-1} \ ,
\label{mz}
\end{eqnarray}
where $g(p)$ and $f(p)$ are the Fourier transform of $g(z)$ and $f(z)$, respectively.
In addition, we have defined
\begin{equation}
\Big({\rho_{n,\vec{p}}^c} \Big)^2 = \Big[ (2 n +1 )\pi  T- i \mu +
\phi_c \Big]^2 + {\vec{p}}\ \! ^2 \ ,
\end{equation}
where the quantities $\phi_c$  are given by the relation $\phi =
{\rm diag}(\phi_r,\phi_g,\phi_b)$. Namely, $\phi_c = c \ \phi_3$
with $c = 1,-1,0$ for $r,g,b$ respectively.
At this stage we need to specify the explicit form of the Polyakov
loop effective potential. Here, we used the fit to QCD lattice
results proposed in Ref.~\cite{Roessner:2006}.

$\Omega^{\rm MFA}$ turns out to be divergent and, thus, needs to
be regularized. For this purpose we use the same prescription as
in Ref.~\cite{Tum:2005}. Namely
\begin{equation}
\Omega^{\rm MFA}_{(reg)} = \Omega^{\rm MFA} - \Omega^{free} +
\Omega^{free}_{(reg)} + \Omega_0 \ , \label{omegareg}
\end{equation}
where $\Omega^{free}$ is obtained from the first term in
Eq.(\ref{granp_reg}) by setting
$\bar \sigma_1 = \bar \sigma_2=0$ and $\Omega^{free}_{(reg)}$ is
the regularized expression for the quark thermodynamical potential
in the absence of fermion interactions,
\begin{equation}
\Omega^{free}_{(reg)} = -4\ T \int \frac{d^3 \vec{p}}{(2\pi)^3}\;
\sum_{c,k}\
\ln\left[ 1 + e^{-\left( \sqrt{\vec{p}^2+m^2}- k \mu + i \phi_c \right)/T} \right] \ ,
\label{freeomegareg}
\end{equation}
with $k=\pm 1$.
Finally, note that in Eq.(\ref{omegareg}) we have included a
constant $\Omega_0$ which is fixed by the condition that
$\Omega^{\rm MFA}_{(reg)}$ vanishes at $T=\mu=0$.

The mean field values $\bar\sigma_{1,2}$ and $\bar{\phi}_3$ at a
given temperature or chemical potential, are obtained from a set
of three coupled ``gap'' equations. This set of equations follows
from the minimization of the regularized thermodynamical
potential, that is
\begin{equation}
\frac{\partial\Omega_{\rm MFA}^{reg}}{\partial\bar\sigma_{1}} =
\frac{\partial\Omega_{\rm MFA}^{reg}}{\partial\bar\sigma_{2}}=
\frac{\partial\Omega_{\rm MFA}^{reg}}{\partial\bar{\phi}_3}=0 \ .
\label{fullgeq}
\end{equation}
Once the mean field values are obtained, the $(T,\mu)$ behavior of other
relevant quantities can be determined.

In order to fully specify the model under consideration we have to
fix the model parameters as well as the form factors $g(q)$ and
$f(q)$ which characterize the non-local interactions. Following
Ref.\cite{Noguera:2008}, we consider two different type
of functional dependencies for these form factors. The first one
corresponds to the often used exponential forms,
\begin{equation}
g(q)= \mbox{exp}[-q^{2}/\Lambda_{0}^{2}] \qquad , \qquad f(q)=
\mbox{exp}[-q^{2}/\Lambda_{1}^{2}] \ . \label{regulators}
\end{equation}
Note that the range (in momentum space) of the nonlocality in each
channel is determined by the parameters $\Lambda_0$ and
$\Lambda_1$, respectively. Fixing the $T=\mu=0$ values  of $m_c$
and chiral quark condensate to reasonable values $m_c = 5.7$ MeV
and $\langle\bar{q}q\rangle^{1/3} = 240$ MeV the rest of the
parameters are determined so as to reproduce the empirical values
$f_\pi = 92.4$ MeV and $m_\pi = 139$ MeV, and $Z(0) = 0.7$ which
is within the range of values suggested by recent lattice
calculations\cite{Parappilly:2005}. In what follows this choice of
model parameters and form factors will be referred as
parametrization S1. The second type of form factor functional
forms we consider is given by
\begin{eqnarray}
\!\!\!\!\!\!\!\!\!
g(q)  = \frac{1+\alpha_z}{1+\alpha_z\ f_z(q)} \frac{\alpha_m \ f_m (q) -
m_{f}\ \alpha_z f_z(q)} {\alpha_m - m_f \ \alpha_z }
\quad , \quad
f(q)  = \frac{ 1+ \alpha_z}{1+\alpha_z \ f_z(q)} f_z(q) \ ,
\label{regulators_set2}
\end{eqnarray}
where
\begin{equation}
f_{m}(q) = \left[ 1+ \left( q^{2}/\Lambda_{0}^{2}\right)^{3/2}
\right]^{-1} \qquad ; \qquad f_{z}(q) = \left[ 1+ \left(
q^{2}/\Lambda_{1}^{2}\right) \right]^{-5/2} \ .
\label{parametrization_set2}
\end{equation}
As shown in Ref.\cite{Noguera:2008}, with a convenient choice of
parameters one can very well reproduce the momentum dependence of
mass and the renormalization function obtained in a Landau gauge
lattice calculation as well as the physical values of $m_\pi$ and
$f_\pi$. In what follows this parametrization will be referred as
S2. Finally, in order to compare with previous studies where the
wavefunction renormalization of the quark propagator has been
ignored we consider a third parametrization (S3). In such case we
take $Z(p)$ = 1 (setting $f(p)$ = 0) and exponential
parametrization for $g(p)$. The values of the model parameters for
each of the chosen parameterizations are summarized in Table I.

\begin{table}[h]
\caption{Sets of parameters.}
{\begin{tabular} [c]{ccccc}\hline &
& \hspace{.5cm} S1 \hspace{.5cm} & \hspace{.5cm} S2 \hspace{.5cm}
& S3 \\\hline
$m_{c}$ & MeV & 5.70 & 2.37 & 5.78\\
$G_{s} \Lambda_{0}^{2}$ &  & 32.030 & 20.818 & 20.650\\
$\Lambda_{0}$ & MeV & 814.42 & 850.00 & 752.20\\
$\kappa_{P}$ & GeV & 4.180 & 6.034 & $-$\\
$\Lambda_{1}$ & MeV & 1034.5 & 1400.0 & $-$\\\hline
\end{tabular}
\label{tab1}}
\end{table}

\section{Results}

We start by analyzing the behavior of some mean field quantities
as functions of $T$ and $\mu$. Since the results obtained for our
three different parameterizations are qualitatively quite similar
we only present explicitly those corresponding to the
parametrization S1.
\begin{figure}[hbt]
\includegraphics[width=\textwidth]{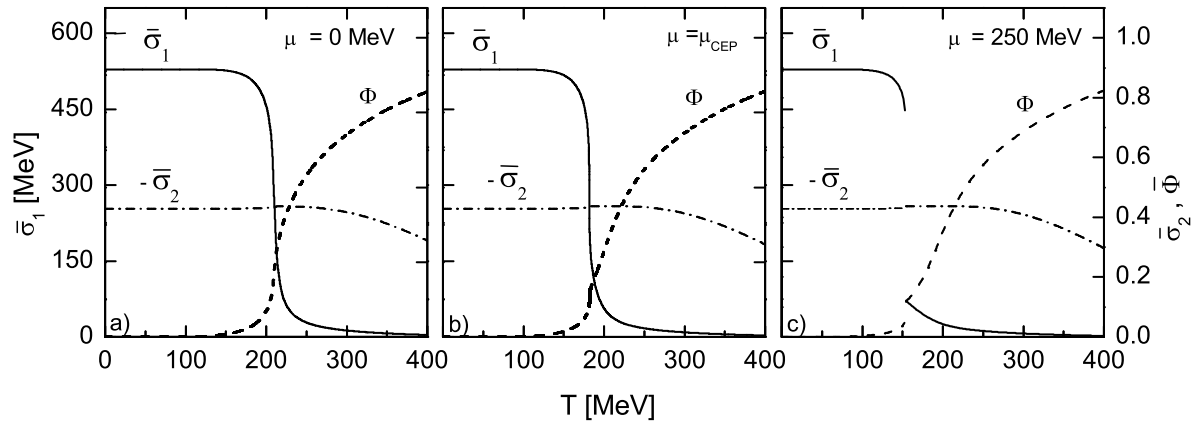}
\caption{Mean fields $\bar \sigma_1$, $\bar \sigma_2$ and $\bar
\Phi$ as functions of $T$ for low (left), high (right) and CEP
(central) chemical potentials. Note that the scale to the left
corresponds to that of $\bar \sigma_1$ while that to the right to
$\bar \sigma_2$ and $\bar \Phi$. Since $\bar \sigma_2$ turns out
to be negative we plot $- \bar \sigma_2$.} \label{s1}
\end{figure}
They are given in Fig.1 where we plot $\bar
\sigma_1$ , $\bar \sigma_2$ and $\bar \Phi$ as functions of $T$
for some values of the chemical potential. Fig. 1a shows that at
$\mu = 0$ there is a certain value of $T$ at which $\bar \sigma_1$
drops rapidly signalling the existence of a chiral symmetry
restoration crossover transition, its position being determined by
the peak of the chiral susceptibility. At basically the same
temperature the Polyakov loop $\bar \Phi$ increases which can be
interpreted as the onset of the deconfinement transition. As $\mu$
increases there is a certain value of $\mu=\mu_{CEP}$ above which
the transition starts to be discontinuous. At this precise
chemical potential the transition is of second order. This
situation is illustrated in Fig.1b. The corresponding values
$(T_{CEP}, \mu_{CEP})$ define the position of the so-called
``critical end point''. As displays in Fig.1c, for $\mu >
\mu_{CEP}$ the transition becomes discontinuous, i.e. of first
order. Finally, for chemical potentials above $\mu_c(T=0)\simeq
310$ MeV the system is in the chirally restored phase for all
values of the temperature. It is important to note that although
$\bar \sigma_2$ appears to be rather constant in Fig.1, at higher
values of $T$ it does go to zero as expected. Concerning the
deconfinement transition we see that as $\mu$ increases there
appears a region where system remains in its confined phase
(signalled by $\bar \Phi$ smaller than $\simeq 0.3$) even though
chiral symmetry has been restored. This corresponds to the
recently proposed quarkyonic phase\cite{McLerran:2007qj}.

\begin{figure}[hbt]
\includegraphics[scale=0.6]{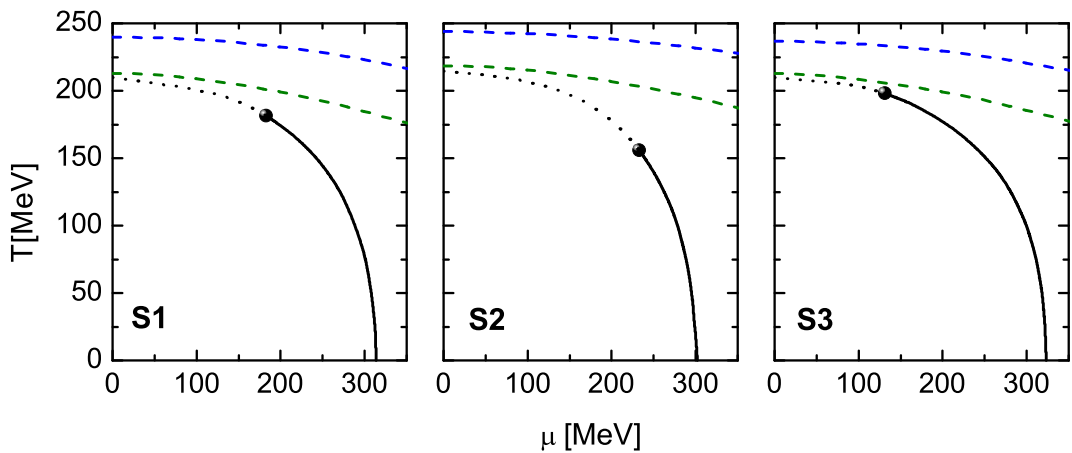}
\caption{Phase diagrams for the three parameterizations
considered. S1 and S2 include quark wave function renormalization
while S3 does not. S1 and S3 correspond to exponential form
factors while S2 to lattice motivated form factors. The dotted
line corresponds to the line of crossover chiral transition and
the full line to that of first order chiral transition. The dashed
lines correspond to the deconfinement transition (the lower and
upper lines being for $\bar \Phi = 0.3$ and $\bar \Phi = 0.5$,
respectively).} \label{phase diagrams} \end{figure}

The phase diagrams corresponding to our three different
parameterizations are given in Fig.\ref{phase diagrams}. Here the
dotted line corresponds to the line of crossover chiral transition
while the full line to the line of first order chiral transition.
The dashed lines correspond to the deconfinement transition (the
lower and upper lines correspond to $\bar \Phi = 0.3$ and $\bar
\Phi = 0.5$, respectively). Comparing those of S1 and S3 we see
that the main effect of the wave function renormalization term is
to shift the location of the CEP towards lower values of $T$ and
higher values $\mu$. Concerning the lattice adjusted
parametrization S2 we observe that it leads to even lower values
of $T_{CEP}$ and higher values $\mu_{CEP}$.

\section{Summary and conclusions}

A non-local extension of the PNJL model momentum which leads to
momentum dependent quark mass and wave function renormalization
has been studied. This model provides a simultaneous description
for the deconfinement and chiral phase transition. Non-local
interactions have been described by considering both a set of
exponential form factors, and a set of form factors obtained from
a fit to the mass and renormalization functions obtained in
lattice calculations. The resulting phase diagrams turn out to be
qualitative similar, the position of the critical end point being
the feature which depends more crucially on each particular
parametrization.

\section*{Acknowledgments}

We would like to thank the members of the Organizing Committee for their warm hospitality
during the workshop. This work has been supported in part by ANPCyT (Argentina),
under grant PICT07 03-00818.

\end{document}